\documentclass{iau_FM}
\usepackage[dvipdfmx]{graphicx}

\newcommand       \mum          {\,{\rm \mu m}}
\newcommand     \simlt  {\lower.5ex\hbox{$\buildrel < \over \sim$}}
\newcommand     \simgt  {\lower.5ex\hbox{$\buildrel > \over \sim$}}
\newcommand     \ltsim  {\lower.5ex\hbox{$\buildrel < \over \sim$}}
\newcommand     \gtsim  {\lower.5ex\hbox{$\buildrel > \over \sim$}}

\title[Nano dust in space and astrophysics] 
{Nano dust in space and astrophysics}

\author[Mann, Li \& Tanaka]  
{Ingrid Mann$^1$, Aigen Li$^{2}$, 
 \and Kyoko K. Tanaka$^3$}

\affiliation{$^1$
UiT The Arctic University of Norway, Troms\o, Norway 
\\ email: {\tt ingrid.b.mann@uit.no} 
\\[\affilskip]
$^2$University of Missouri, Columbia, Missouri, USA
 \\email: {\tt lia@missouri.edu} 
\\[\affilskip]
$^3$Tohoku University, Sendai, Japan 
 \\email: {\tt kktanaka@astr.tohoku.ac.jp}}

\pubyear{2018}
\jname{Astronomy in Focus, Volume 2} 
\editors{Maria Teresa V.T. Lago, ed.}

\begin{document}
\maketitle

\begin{abstract} 
The theme of this focus meeting is related to the detection, 
characterization and modeling of nano particles ---
cosmic dust of sizes of roughly 1 to 100 nm ---
in space environments like the interstellar medium, 
planetary debris disks, the heliosphere,
the vicinity of the Sun and planetary atmospheres, 
and the space near Earth. 
Discussions  focus on nano dust that forms
from condensations and collisions and from
planetary objects, as well as its interactions 
with space plasmas like the solar and stellar winds, 
atmospheres and magnetospheres. 
A particular goal is to bring together 
space scientists, astronomers, astrophysicists, 
and laboratory experimentalists 
and combine their knowledge 
to reach cross fertilization 
of different disciplines.  
\end{abstract}

Nano dust particles are intermediate between molecules and bulk
matter and are widespread in space 
(see Li \& Mann 2012, Xie et al.\ 2018). 
Because of their finite small size and large surface-to-volume
ratio, the physical properties of nano grains are often peculiar, 
being qualitatively different from those of bulk materials. 
Different behavior is found, e.g., in heat capacity, 
melting temperature, surface energy, diffusion coefficient, 
and optical properties 
(e.g., see Halperin 1986).
Especially, clusters of $\sim$\,1--10\,nm 
are expected to reveal strongly variable size-dependent properties 
such as electronic structure, binding energy and dielectric function 
which determine how they interact with gas particles and the
electromagnetic radiation.
Larger clusters, with many thousands of atoms 
and diameters in the range of 10\,nm and more,
have a behavior smoothly varying with size 
and approaching bulk properties as size increases.

\begin{figure}
\centering
\hspace*{-12mm}
\includegraphics[width=10.8cm]{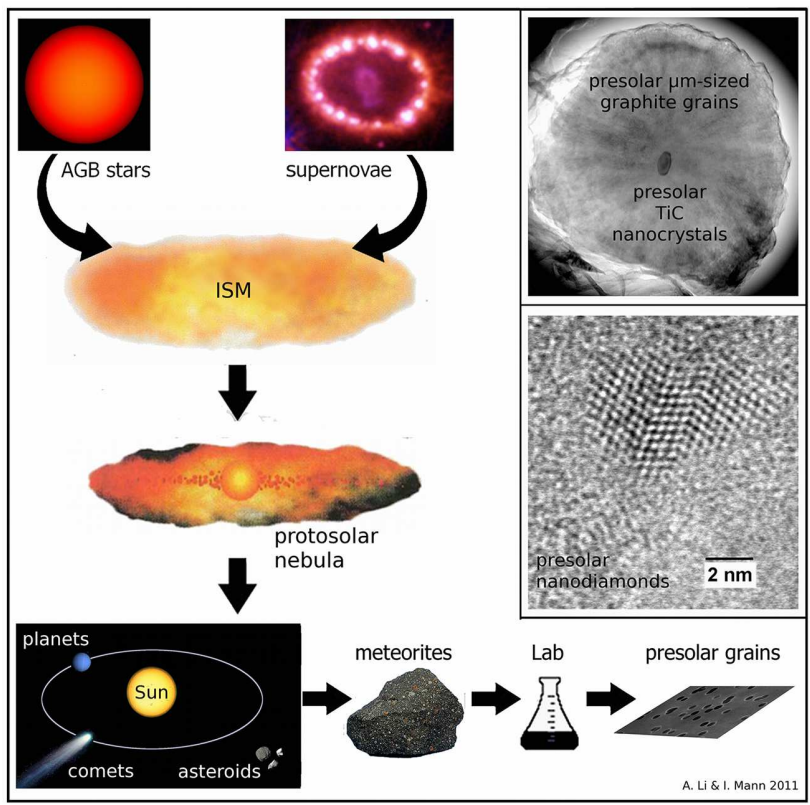}
\caption{
         \label{fig:presolar_grains}
         Schematic illustration of the history 
         of cosmic dust grains,
         from their condensation in 
         stellar winds of
         asymptotic giant branch (AGB) stars or
         in supernova ejecta,
         to their injection into the ISM,
         and subsequent incorporation into 
         the dense molecular cloud 
         from which our solar system formed
         (i.e., protosolar nebula).
         These grains survived all the violent processes 
         occurring in the ISM (e.g., sputtering by shock waves)
         and in the early stages of solar system formation 
         and were incorporated into meteorite parent bodies.
         Finally, they were collisionally liberated from their   
         parent bodies and entered the Earth atmosphere,
         making them available for 
         experimental studies in terrestrial laboratories,
         and therefore allowing one to separate them from
        the meteorite or interplanetary dust material 
        in which they are embedded.         
	 Inserted are the TEM 
         ({\it Transmission Electron Microscopy}) 
         images of presolar nanodiamond grains
         (Banhart et al.\ 1998)
         and a presolar TiC nanocrystal within 
         a micrometer-sized presolar graphite spherule.
         (Bernatowicz et al.\ 1996). 
         }
\end{figure}

While the exact role of nano dust is not fully understood yet, 
those nanoclusters should play an important role, 
since, because of their large surface area 
(relative to their small mass), they interact more efficiently 
with particles and fields. Interstellar nano dust grains,
probably generated by fragmentation 
of larger carbonaceous grains 
(e.g., see Onaka et al.\ 2018),
dominate the far ultraviolet extinction 
at $\lambda^{-1}\gtsim6\mum^{-1}$
as well as the near- and mid-infrared emission 
at $\lambda\simlt 60\mum$
of the interstellar medium (ISM) of the Milky Way and
external galaxies (see Li 2004).
The ``anomalous microwave emission'', 
an important Galactic foreground of the cosmic microwave
background radiation in the $\sim$\,10--100\,GHz region, 
suggest that the ISM contains a considerable amount 
of nano grains (see Dickinson et al.\ 2018), 
possibly of polycyclic aromatic hydrocarbon (PAH)
(Draine \& Lazarian 1998) 
or silicate (Hoang et al.\ 2016, Hensley \& Draine 2017)
in composition.
A wide variety of astrophysical environments 
such as reflection nebulae, planetary nebulae,
and H\,II regions also exhibit a broad, featureless
band at $\sim$\,5400--9500\,${\rm \AA}$
known as the ``extended red emission'', 
generally attributed to photoluminescence 
by nanoparticles (Witt 2000). 
The heating of the interstellar gas 
(Bakes \& Tielens 1994, Weingartner \& Draine 2001)
and the surface layers of protoplanetary disks 
(Kamp \& Dullemond 2004) 
are dominated by nano grains 
through the photoelectrons provided by them. 
The presence of charged nano dust likewise 
influences other space plasma, also leading to 
dusty plasma effects, like waves and instabilities
(see Mann et al.\ 2014).
Nano-sized (or smaller) PAH molecules 
and their ions (e.g., see Peeters 2014),
nano diamonds (Guillois et al.\ 1999,
van Kerckhoven et al.\ 2002),
C60 and C70 as well as their ions 
(Cami et al.\ 2010, Sellgren et al.\ 2010, 
Bern\'e et al.\ 2013, Strelnikov et al.\ 2015),
and possibly also graphene C$_{24}$
(Garc{\'{\i}}a-Hern{\'a}ndez et al.\ 2011, Chen et al.\ 2017)
reveal their presence in astrophysical regions 
through their characteristic 
vibrational spectral features. 
Nanodiamonds (Lewis et al.\ 1989) 
and nano TiC crystals (Bernatowicz et al.\ 1996)
have also been identified as presolar grains 
in primitive meteorites through their isotope anomalies.
Their path from formation in the late stages of stellar
evolution to identification in the laboratory is sketched 
in Figure~\ref{fig:presolar_grains}.

For many years nano dust has been detected 
with in-situ instruments from spacecraft 
in different regions of the solar system
(e.g., see Jones 2012, Hsu et al.\ 2012,
Meyer-Vernet \& Zaslavsky 2012).
In the inner heliosphere of our solar system, 
nano dust forms from the dust-dust collisions 
in the zodiacal cloud (see Mann \& Czechowski 2012)
and from sun-grazing comets 
(e.g., see Ip \& Yan 2012, Mann 2017).
The interaction of nanodust 
with the solar radiation
and with gas particles 
often differs from that of 
the larger solar system dust	
(see Mann et al.\ 2014).
A notable recent finding is that nano dust 
in the heliosphere is deflected and
accelerated in the solar wind 
(Mann et al.\ 2007, Czechowski \& Mann 2010,
Juhasz \& Horanyi 2013, Zaslavsky 2015).
Astronomical observations suggest 
the presence of nano dust also in 
circum-stellar debris disks around 
main-sequence stars under conditions 
similar to the inner heliosphere.
Nanodust is suggested to explain 
the hot dust component in circumstellar 
debris disks around main-sequence stars
(e.g., see Lebreton et al.\ 2013, Su et al.\ 2013)
which, due to dynamics effects, 
possibly stays near the star for prolonged time 
(Kobayashi et al.\ 2008, Rieke et al.\ 2016). 
The conditions in the inner debris disks 
were recently systematically studied 
for a number of stars 
(Kirchschlager et al.\ 2017, 
Kimura et al.\ 2018).
In-situ measurements from sounding rockets 
detect nano dust in the Earth's upper atmosphere 
(mesosphere), where it forms from the re-condensation 
of metallic compounds 
produced from ablating meteoroids
(cf. Plane et al.\ 2017).
This dust --- termed meteoric smoke --- provides 
condensation nuclei for noctilucent clouds 
(first reported in 1886, and almost certainly 
a harbinger of climate change 
in the upper atmosphere). 
Meteoric smoke is also implicated 
in the formation and freezing of stratospheric clouds
(which cause polar ozone depletion) 
and in the chemistry of clouds 
and atmospheres of, 
e.g., Mars, Venus, and Titan
(Plane et al.\ 2018).
Nano dust is probably also observed in comets
(Utterback \& Kissel 1990).
Planetary vulcanic plumes 
and impacts on planetary objects are sources of
interplanetary nano dust as, e.g., observed near 
the surface of the Moon (Wooden et al.\ 2016).

Nanodust interacts efficiently with particles 
and fields and in plasmas. The vast majority 
of our universe is plasma in which the heavy
chemical elements are often contained in small solid dust particles
that carry electric surface charge. A large fraction of the plasma is
therefore dusty plasma, where dust participates in and gives rise to
charge collective effects. 
Examples for dusty plasma 
are the ISM, the Earth's ionosphere, 
the ring systems of planets 
as well as the surface layers of moons 
and in general of solar system objects 
that are not surrounded by an atmosphere
(see Mendis \& Rosenberg 1994).
Although dusty plasma is extensively studied, 
only a few observations in space
are fully described with existing theory.

The research on the dynamics of nano dust 
in the heliosphere at present progresses 
motivated by the detection with space instruments 
on several different spacecraft.
We expect a wealth of new observations 
in the near future. NASA has launched 
the {\it Parker Solar Probe} in August 2018 
and ESA will soon launch the {\it Solar Orbiter}. 
Both spacecraft explore the most inner heliosphere 
and plasma in the vicinity of the Sun,
including a region where we expect 
nanodust is being formed.
Nanodust is possibly detected 
with plasma wave measurements 
when it impacts the spacecraft 
(Bale et al.\ 2016).
Observing the near- and mid-infrared emission 
of the ISM which is dominated 
by nano dust is one focus of
{\it JWST} which will be launched in 2021.
Laboratory astrophysics is today 
a well-established field 
and in recent years further progress 
on dust studies was made with sample returns,
from which the knowledge, 
often referring to larger dust, 
still provides information on nano dust.
The $\sim$\,8--13\,$\mu$m spectra 
recently measured in situ for free-flying 
nucleating nano silicate particles were 
shown to be remarkably consistent with 
that observed in oxygen-rich evolved stars
(Ishizuka et al. 2015).
Recent microgravity experiments 
using sounding rockets also provided us 
with useful information on dust 
nucleation and growth 
(e.g., see Kimura et al.\ 2017).

An important development was recently made
through numerical simulations of dust growth. 
Molecular dynamics models are applied to study 
the vapor-to-solid transition 
(e.g., see Tanaka et al.\ 2017). 
Quantum chemical calculations of large molecules
also help to study the link between molecules 
and nanodust (e.g., see Kwok \& Zhang 2011). 
To simulate particle growth, dynamics combined 
with clustering is simulated for conditions of,
e.g., molecular clouds (Mattsson 2016).

During the General Assembly in Vienna,  
the Focus Meeting 
``{\it Nano Dust in Space and Astrophysics}'' 
(FM10) on 28--29 August 2018
has brought together space physicists 
who study nano dust in the heliosphere 
and specialists from physics, astrophysics, 
as well as atmospheric research to make progress 
in understanding nano dust particles by combining 
their knowledge on dust under a wide range of
space conditions. Knowledge on nano dust
is also gained from studies of larger particles 
that progressed through laboratory astrophysics
and analysis of returned samples.

We thank all our colleagues who participated 
in the meeting and the members of 
the science committee 
who participated in preparing the programme: 
Alexander G.G.M. Tielens (The Netherlands), 
Anja C. Andersen (Denmark), 
Anny-Chantal Levasseur-Regourd (France), 
Biwei Jiang (China), 
Chris M. Wright (Australia), Farid Salama (USA), 
John Plane (United Kingdom), Joseph A. Nuth (USA), 
Khare Avinash (India), Sun Kwok (Hong Kong, China), 
Thomas Pino (France) and Veronica Motta (Chile).


\end{document}